\documentclass[useAMS]{mn2e}
\usepackage[dvips]{graphicx}
\usepackage{amsmath,amssymb}
\title[TeV blazar variability: the firehose instability?]{TeV blazar variability: the firehose instability?}
\author[Subramanian, Shukla and Becker]{Prasad Subramanian$^{1}$\thanks{E-mail: p.subramanian@iiserpune.ac.in}, Amit Shukla$^{2}$, Peter A Becker$^{3}$\\
$^{1}$Indian Institute of Science Education and Research, Garware Circle, Pashan, Pune - 411021, India\\
$^{2}$Indian Institute of Astrophysics, Koramangala, Bangalore - 560034, India\\
$^{3}$College of Science, George Mason University, Fairfax, VA 22030, USA}
\begin{document}

\date{}

\pagerange{\pageref{firstpage}--\pageref{lastpage}} \pubyear{2011}

\maketitle

\label{firstpage}

\begin{abstract}
Recently observed minute timescale variability of blazar emission at TeV energies has imposed severe constraints on jet models and TeV emission mechanisms.
%The TeV emission is assumed to be due to inverse Compton emission off the electrons, which have a random Lorentz factor $\gamma \sim 10^6$. 
We focus on a robust jet instability to explain this variability. As a consequence of the bulk outflow of the jet plasma, the pressure is likely to be anisotropic, with the parallel pressure $P_{||}$ in the forward jet direction exceeding the perpendicular pressure $P_{\perp}$. Under these circumstances, the jet is susceptible to the firehose instability, which can cause disruptions in the large scale jet structure and result in variability of the observed radiation.
%We investigate if the large-scale structure of the jet can be disbrupted by the firehose instability. 
For a realistics range of parameters, we find that the growth timescale of the firehose instability is $\approx$ a few minutes, in good agreement with the observed TeV variability timescales for Mrk 501 (Albert et al. 2007) and PKS 2155-304 (Aharonian et al. 2007). 
%We therefore conclude that the relativistic firehose instability is a viable candidate for explaining minute timescale TeV variability.  
\end{abstract}

\begin{keywords}
keywords
\end{keywords}

\section{Introduction}
Several blazars have recently displayed minute-timescale variability at TeV energies (Albert et al. 2007; Aharonian et al. 2007). This has resulted in a great deal of interest in the emission mechanisms responsible for these TeV flares. Suggestions range from a coherent instability in a compact emission region (e.g., Begelman, Fabian \& Rees 2008), misaligned minijets inside the main jet (e.g., Giannios, Uzdensky \& Begelman 2010), jet deceleration (Georganoupolos \& Kazanas 2003; Levinson 2007) to wiggles in an anisotropic electron beam directed along the jet (Ghisellini et al. 2009). Considerable attention has also been paid to correlated variability at other wavelengths (such as optical wavelengths) exhibited by TeV blazars (e.g., Gopal-Krishna et al. 2011 and references therein). In general, there have been several mechanisms proposed for producing the observed variability in the jet emission, ranging from plasma mechanisms (Krishan \& Wiita 1994) to beamed radiation (e.g., Crusius-W\"atzel \& Lesch 1998). TeV variablity is somewhat unique, in that explaining even steady-state TeV emission from blazars poses significant challenges. Electron emission (via the synchroton self-compton process or the external inverse Compton process) is frequently invoked. While electrons are certainly the readiest radiators, both these mechanisms require very highly relativistic electrons (random Lorentz factors \textbf{$\gamma \approx 10^{4}-10^{5}$} or more). This in turn poses severe electron reacceleration problems. %Protons, on the other hand, do not radiate as readily, and reacceleration problems are therefore not an issue. However, there are few models that can explain mildly relativistic proton jets (e.g., Subramanian, Becker \& Kazanas 1999, BECKER \& LE). Nontheless, it is hard to imagine processes that can accelerate protons to high enough energies so that they can produce TeV emission, either via the $p-p$ process or the photomeson process.
\section[]{Current models for TeV flares: the role of bulk outflow} 
We now pay attention to the role of bulk plasma outflow in a couple of popular models for TeV flares. In the first model we discuss, the bulk plasma outflow is necessary to alleviate the problem of copius pair production (and consequent degradation of TeV photons) in the TeV emission region. In the second one, the idea of a highly anisotropic, directed beam of TeV emitting electrons is central to explaining the observed variability. We point out later that the bulk Lorentz factor ($\Gamma$) of the plasma outflow is related to pressure anisotropy in the jet. The magnitude of $\Gamma$ (and therefore that of the pressure anisotropy) is limited by the excitation of hydromagnetic waves. We will then examine the role of the pressure anisotropy in exciting the well studied firehose instability in the jet. Jet disruption due to this instability is found to be a viable explanation for the observed variability at TeV energies.
\subsection{TeV variability due to a compact emission region: the pair production problem}
The first scenario we examine is due to Begelman, Fabian \& Rees (2008). In this scenario, the TeV photons are produced via inverse Compton upscattering of an existing soft photon population by energetic electrons. Using a peak frequency of $10^{16}$ Hz for the soft photons and a bulk Lorentz factor of $\Gamma \approx 50$, they estimate that the random Lorentz factor of the energetic electrons $\gamma \approx 10^{4}$ in order to produce TeV emission. They relate the short variability timescales (of the order of a few minutes) to the size of the emission region via the light crossing time argument. This argument implies that the emission originates from a region whose size is a small fraction of the black hole's Schwarzschild radius (Begelman, Fabian \& Rees 2008). For relativistic, Poynting flux dominated jets, it is possible that structures as small as the gravitational radius of the black hole are imprinted on the jet as it is launched, and modulate the emission far away from the central region (Kirk \& Mochol 2010). Another possibility is that the TeV flares could originate in highly localized fluctuations in the outflowing jet plasma due to instabilities.

Importantly, Begelman, Fabian \& Rees (2008) show that the outflowing plasma must have a substantial bulk Lorentz factor $\Gamma \approx 50$ in order for the TeV photons to escape from the compact emission region without producing pairs. If this was not so, copius pair production due to the interaction between the TeV photons and locally produced synchrotron photons would substantially degrade the TeV photons produced. Furthermore, Begelman, Fabian \& Rees (2008) argue that the TeV emitting region needs to be located far away from the central black hole (at least 100 Schwarzschild radii) in order to avoid rapid degradation of the TeV photons due to pair production on the soft photon background of the accretion disk (e.g., Becker \& Kafatos 1995). Assuming a reasonable jet opening angle, this implies that the size of the emission region is around two orders of magnitude smaller than the transverse dimension of the jet, at the distance where the variable TeV emission originates. This scenario thus envisages a very compact emission region comprising energetic electrons flowing out with a bulk Lorentz factor $\approx 50$.

\subsection{TeV variability due to beamed electron distribution?}
Ghisellini et al. (2009) outline an interesting scenario where the TeV emission is due to a highly anisotropic electron beam. The electrons are almost co-aligned in a narrow beam. Another way of stating this could be to say that the electrons are hot along the beaming direction and cold in the perpendicular direction; the jet pressure (as observed by a distant `lab'' observer) is highly anisotropic. It would be possible to define a meaningful bulk Lorentz factor along the beaming direction, but the jet pressure would be anisotropic even in the bulk comoving frame. The electrons could be emitting TeV radiation via the inverse Compton scattering of soft photons. As an extreme case, they calculate the electron Lorentz factor in a situation where there is no bulk motion and the soft photons are isotropically distributed (although, as mentioned earlier, the highly anisotropic electron beam clearly implies ordered bulk motion). 
%They envisage a scenario where there is no bulk motion of the energetic electrons, and the soft photons are isotropically distributed. 
They find that the electron random Lorentz factor $\gamma$ needs to be $\approx 10^{6}$ in order to account for TeV photons produced via inverse Compton scattering off soft photons of energies $\sim 1$ eV. Ghisellini et al. (2009) envisage large-scale wiggles in the beam, so that the radiation would be observed only for the fraction of time that the beam points toward the observer, leading to variability. The electron beam is assumed to retain its coherence, and the variability is only because of the fact that the large scale magnetic field points toward the observer for a limited time. Such a scenario avoids explicit reference to a compact emission region.

A central feature of Ghisellini et al. (2009)'s model is the highly beamed, anisotropic electron distribution, which leads to the parallel pressure greatly exceeding the perpendicular one. We show below that the electron pressure anisotropy is unlikely to assume extreme values. We show, however, that even modest amounts of pressure anisotropy are enough to destabilize the large-scale structure of the jet via the firehose instability, and this could result in observed variability.

\section{Our scenario}
Our scenario is similar to Begelman et al. (2008)'s one in that the radiating electrons possess a random Lorentz factor $\gamma$ ranging from $10^4$ to $10^6$ as well as a bulk Lorentz factor $\Gamma$. The issue of pressure anisotropy highlighted by Ghisellini et al. (2009) is important in our scenario too.

In this work, we envisage a jet that comprises an electron-proton plasma which streams along the forward jet direction. All the quantities referred to from here onwards are defined in the distant observer's frame of reference. Electrons would be bound to the protons, and stream along with them, but would also possess random Lorentz factors $\gamma \approx 10^4$--$10^6$. Since the electrons have such high random Lorentz factors, and the protons are most likely much colder, the electrons are the dominant contributors to the pressure.
The velocity with which the plasma streams along the jet is typically restricted to the Alfv\'en speed $v_{\rm A}$ via a self-limiting process. This limit on the streaming speed is imposed by hydromagnetic waves emitted by the electrons (Wentzel 1969; Melrose 1970), which also scatter the electrons and limit the degree of anisotropy.
Following Wentzel (1968; 1969) we envisage the equilibrium particle distribution function in momentum ${\mathbf p}$ of the jet plasma to be of the form
\begin{equation}
f({\mathbf p}) = f_{i}({\mathbf p}) \biggl ( 1 + 3\,|\mu| \frac{v_{A}}{c} \biggr) \, ,
\label{eq1aa}
\end{equation}
where $f_{i}({\mathbf p})$ is the isotropic part of the distribution function and satisfies 
\begin{equation}
\int f_{i}({\mathbf p}) d^{3} {\mathbf p} = 1\, .
\label{eq1ab}
\end{equation}
The quantity $\mu$ is the cosine of the particle pitch angle and represents the fraction of the particle momentum along the direction of the bulk motion of the jet. The quantity $c$ is the speed of light. 
%The isotropic part of the distribution function satisfies $\int f_{i} d^{3} p = 1$, where $p$ is the particle momentum. 
The second term in the brackets expresses the anisotropy of particle distribution. 
It may be noted that the anisotropy in this formulation is typically only a perturbation; in other words, we require that $v_{\rm A}/c \ll 1$. Wentzel's (1968) formulation includes a higher order term $\propto \mu^{2}$ on the right hand side of Eq~\ref{eq1aa}, but we don't consider such a term in the interest of simplicity. While we do not appeal to a specific mechanism to generate the electron anisotropy, such as the pitch angle dependence of synchrotron losses (e.g., Melrose 1970; Wentzel 1969) or the intrinsic jet launching mechanism (e.g., Aharonian, Timokhin \& Plyasheshnikov 2002), we emphasize that this distribution (Eq~\ref{eq1aa}) is only mildly anisotropic, and the degree of anisotropy in  is restricted to $\approx v_{\rm A}/c \ll 1$. Furthermore, the bulk streaming speed of the jet is restricted to be approximately equal to the Alfv\'en speed $v_{\rm A}$. 
The ratio of the parallel pressure $P_{||}$ to the perpendicular pressure $P_{\perp}$ in the jet arising from the distribution given by Eq~\ref{eq1aa} is (Wentzel 1968)

\begin{equation}
\frac{P_{||}}{P_{\perp}} = \frac{1 + v_{\rm A}/c}{1 - v_{\rm A}/c} \,
\label{eq1ab}
\end{equation}

Since the bulk streaming speed of the jet is $\approx v_{\rm A}$ , we can write the following expression for the jet bulk Lorentz factor $\Gamma$:

\begin{equation}
\Gamma = (1 - v_{\rm A}^{2}/c^{2})^{-1/2}\, ,
\label{eq1ac}
\end{equation}
keeping in mind the fact that $\Gamma$ cannot greatly exceed unity, since $v_{\rm A}/c \ll 1$.

Before we move on to analyzing the firehose instability arising out of the anisotropic pressure distribution in the jet, we note that electron reacceleration within the jet is essential, in view of the copious radiative losses they experience. The electrons could be energized by resonating with the very hydromagnetic waves they shed, much in the same manner as cosmic rays are thought to be accelerated (e.g., Bell \& Lucek 2000), or by wave-particle interactions with a separate, pre-existing wave population (e.g., Eilek 1979). It is also possible that electrons are reaccelerated by the turbulence initiated by a previous episode of the firehose instability; such a scenario is often invoked for electron acceleration in solar flares (e.g., Paesold \& Benz 1999). The protons, on the other hand, are likely to remain substantially colder.

\subsection{The Firehose instability due to $P_{\parallel} > P_{\perp}$}
The well known firehose instability (also often referred to as the gardenhose instability) is typically operative when the parallel pressure $P_{||}$ in the forward jet direction exceeds the perpendicular pressure $P_{\perp}$ transverse to the jet (e.g., Krall \& Trivelpiece 1973). This nonresonant, fluid instability will result in large-scale displacements in the beam, and possible disruption of the large-scale magnetic field in the jet (e.g., Baker et al. 1988), which could result in variability in the observed TeV emission. Resonant instabilties can also be operative, but they typically only result in a redistribution of electron pitch angles, while nonresonant, fluid instabilities disrupt the large-scale magnetic field. 
%We argue that the jet pressure is unlikely to be very anisotropic. 
The arguments of the previous section suggest that the jet pressure is unlikely to be highly anisotropic. We show hereafter that the growth timescales for the firehose instability in a typical TeV emitting jet are in agreement with observed variability timescales, even for moderate pressure anisotropies.

\subsection{Growth timescale of the firehose instability}
While there are several treatments of the firehose instability, we concentrate on one that offers a convenient analytical approximation to the maximum growth rate of the relativistic firehose instability from a solution of the dispersion equation. The maximum growth rate $\omega_{\rm gr}$ of this instability can be approximated to within $\approx$ 20\% by the following expression: (Noerdlinger \& Yui 1969)

\begin{equation}
\omega_{\rm gr} = 0.55\,\omega_{\rm LR}\,\frac{(1 - H)}{8 + \Lambda^{1/2}}\, .
\label{eq2}
\end{equation}
The quantity $\omega_{\rm LR}$ is the mean relativistic gyrofrequency.

given by
\begin{equation}
\omega_{\rm LR} = \frac{e\, B}{\gamma\,m_{e}\,c}\,\,\,\,\,\,{\rm s^{-1}} ,
\label{eq2a}
\end{equation}
where $e$ is the electron charge, $B$ is the large-scale magnetic field in Gauss, $\gamma$ is the random Lorentz factor of the TeV emitting electrons, $m_{e}$ is the electron mass and $c$ is the speed of light.
The growth timescale given by Eq~(\ref{eq2}) is an approximation to the growth rate derived from solving the dispersion relation for the instability. 

The quantity $H$ is the magnetic pressure nondimensionalized by the particle anisotropy, and is defined as

\begin{equation}
H = \frac{B^{2}}{4 \, \pi \, P_{||}(1 - P_{\perp}/P_{||})} \, .
\label{eq3}
\end{equation}
The quantity $\Lambda$ is defined as

\begin{equation}
\Lambda = 1 + \frac{c^{2}}{v_{\rm A}^{2}}\, ,
\label{eq3a}
\end{equation}
and can be related to the bulk Lorentz factor $\Gamma$ via Eq~(\ref{eq1ac}).
It may be noted that the gyrofrequency is used in Eq~(\ref{eq2}) only as a convenient parametrization for the growth timescale, and does not suggest any kind of resonance; the relativistic firehose instability is a nonresonant one.

Identifying the Alfv\'en speed as 

\begin{equation}
v_{\rm A}^{2} \equiv \frac{B^{2}}{4\,\pi\,m_{p}\,N}\, ,
\label{eq4}
\end{equation}

where $m_{p}$ is the proton mass and $N$ is the particle number density, we can express $H$ (Eq~\ref{eq3}) as

\begin{equation}
H = \frac{m_{p}}{m_{e}}\,\frac{v_{A}^{2}}{\gamma\,c^{2}\,(1 - P_{\perp}/P_{||})} \, ,
\label{eq5}
\end{equation}
where $m_{e}$ denotes the electron mass. We have used the proton mass in computing the Alfv\'en velocity, since they are the heavier species, and are the primary contributors to the matter density. We have also used $P_{||} = \gamma\,N\,m_{e}\,c^{2}$ in deriving Eq~(\ref{eq5}), since we expect the energetic electrons (with $\gamma \approx 10^6$) to be the primary contributors to the jet pressure.

The maximum growth timescale of the relativistic firehose instability is 
\begin{equation}
t_{\rm gr} = 2 \pi/\omega_{\rm gr}\, .
\label{eq6}
\end{equation}
As mentioned in \S~3, the growth timescale is evaluated in the frame of the distant observer. Using Eqs~(\ref{eq5}), (\ref{eq3a}), (\ref{eq2a}), (\ref{eq2}), (\ref{eq1ac}) and (\ref{eq1ab}) in Eq~(\ref{eq6}), we get
{\begin{eqnarray}
\nonumber
t_{\rm gr} = 2 \pi\,\frac{\gamma\,m_{e}\,c}{0.55\,e\,B}\,\biggl [8 + \biggl (1 + (1 - \Gamma^{-2})^{-1} \biggr )^{1/2} \biggr ]\, \times \\
\biggl [ 1 - \frac{1}{\gamma}\,\frac{m_{p}}{m_{e}}(1 - \Gamma^{-2})\,\frac{1 + (1 - \Gamma^{-2})^{1/2}}{2\,(1 - \Gamma^{-2})^{1/2}} \biggr ]^{-1}
\label{eq6a}
\end{eqnarray}}

\section{Results}
\begin{figure}
\includegraphics[]{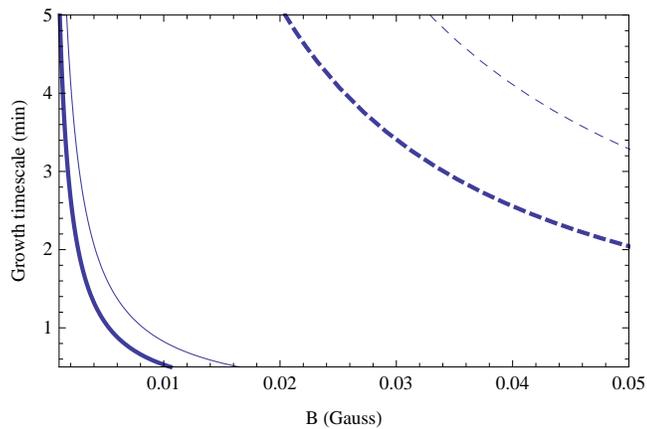}
\caption{The predicted variability timescale $t_{\rm gr}$ in minutes as a function of the magnetic field $B$ in Gauss. The thin solid line is for $\gamma = 5 \times 10^4$, $\Gamma = 1.01$ and the thick solid line is for $\gamma = 5 \times 10^4$, $\Gamma = 5$. The thin dashed line is for $\gamma = 10^6$, $\Gamma = 1.01$ and the thick dashed line is for $\gamma = 10^6$, $\Gamma = 5$.}
\end{figure} 
We show the results for the maximum growth timescale evaluated using Eq~(\ref{eq6a}) in Figure 1. 
We use the bulk Lorentz factor $\Gamma$, the random electron Lorentz factor $\gamma$ and the value of the ambient large scale magnetic field $B$ as our parameters.  
We now comment on the range of values we use for the parameters employed in the calculations. 
\subsection{Range used for $\gamma$}
As mentioned earlier, we assume that the TeV radiation is produced due to energetic electrons upscattering soft photons via the inverse Compton mechanism. Values quoted in the literature for the random Lorentz factor $\gamma$ used to generate TeV emission via the inverse Compton mechanism range from $\approx 10^{4}$ (e.g., Begelman et al 2008) to $\approx 10^{6}$ (Ghisellini et al. 2009). We use two values for the random Lorentz factor of the electrons in our calculations: $\gamma = 5 \times 10^{4}$ and $10^{6}$. 
\subsection{Range of magnetic field values}
We consider magnetic field values in the range $0.001 < B < 0.05$ G. This is representative of the values derived for TeV blazars. Using a one-zone synchrotron self-Compton model, multi-wavelength data fitting requires that the magnetic field in the emission regions of TeV blazars MrK 421 and MrK 501 be $\approx 0.01$ G (Abdo et al. 2011a; 2011b). Similar considerations yield a value for the magnetic field of 0.03 G for the high state of MrK 421 (Shukla et al. 2012). Giroletti et al. (2004) obtain $B \approx 0.01$ G from equipartition considerations in relation to radio observations of the TeV blazar MrK 501. 
\subsection{Range used for $\Gamma$}
We next discuss the values we adopt for the bulk Lorentz factor $\Gamma$. We use $\Gamma = 1.01$ (corresponding to $P_{||}/P_{\perp} = 1.32$) and $\Gamma = 5$ (corresponding to $P_{||}/P_{\perp} = 100$).
Since the assumption regarding mild anisotropy (Eq~\ref{eq1aa}) requires that the ratio of the Alfv\'en speed to the speed of light $v_{\rm A}/c \ll 1$, the bulk Lorentz factor $\Gamma$ needs to be restricted to values that are not appreciably greater than unity (Eq~\ref{eq1ac}). These values for $\Gamma$ imply relatively modest constraints on the jet production mechanism. Although rather high values for $\Gamma$ are often discussed in the literature, it is worth mentioning that multi-epoch radio mapping of parsec-scale jets in TeV emitting blazars suggest that $\Gamma \approx$ 1--3 (Piner, Pant \& Edwards 2008; also see Giroletti et al. 2004). Similar values for $\Gamma$ are also implied by blazar unification schemes (e.g., Urry \& Padovani 1991). Theoretical considerations regarding launching electron-proton jets imply that asymptotic bulk Lorentz factors are likely to be limited to values well below 10 (Subramanian, Becker \& Kazanas 1999).

The broad conclusion from Figure 1 is that a wide range of realistic parameters ($5 \times 10^{4} < \gamma  < 10^{6}$, $1.01 < \Gamma < 5$, $0.001 < B < 0.05$ G) yield growth timescales of the order of a few minutes for the firehose instability.

\section{Summary and Discussion}
We have considered the problem of minute timescale variability in TeV blazars. 
We assume that the TeV radiation is produced by highly relativistic electrons ($\gamma \approx 10^4 -10^6 $) scattering off soft photons via the inverse Compton mechanism. 
The jet is beamed towards the observer with a bulk Lorentz factor $\Gamma$. These aspects are similar to those considered in previous treatments (e.g., Begelman et al. 2008; Ghisellini et al. 2009). However, we point out here that bulk streaming of the jet plasma results in the parallel pressure $P_{||}$ being $\gtrsim$ the perpendicular pressure $P_{\perp}$. We furthermore show that even such a mild pressure anisotropy can give rise to a nonresonant, fluid instability called the firehose instability, which can result in a disruption of the large-scale jet over timescales of a few minutes. In other words, we have shown that the firehose instability is essentially unavoidable in the type of situations considered by Begelman et al. (2008) and Ghisellini et al. (2009). 
%Such a disruption of the large-scale jet features can be manifested as variability in the observed TeV emission.

For $\gamma$ ranging from $5 \times 10^4$ to $10^6$, $\Gamma \gtrsim 1$ and $0.001 < B < 0.05$ G, we find that the firehose instability growth timescale is of the order of a few minutes (Figure 1), which is in agreement with the observed TeV variability timescales for Mrk 501 (Albert et al. 2007) and PKS 2155-304 (Aharonian et al. 2007). Hence the disruption of the large-scale jet features due to the firehose instability provides a natural explanation for the observed variability at TeV energies. 
%This suggests that jet disruption via the relativistic firehose instability is a viable mechanism for explaining TeV blazar variability. 

%Before we end, mention must be made of the kink instability, which is another well studied mode often held responsible for large-scale displacements in the beam. The kink mode is a current driven instability, which relies on the presence of a strong axial current. It is not obvious that TeV emitting jets carry a large net axial current, although the large-scale magnetic field does become predominantly toroidal above the Alfve\'n point, suggesting the presence of some axial current.

%\section*{Acknowledgments}

%We thank --- .

%\begin{figure}
%\includegraphics[width=8.0 cm]{fireh1.eps}
%\caption{The predicted variability timescale $t_{\rm gr}$ in minutes as a function of the magnetic field $B$ in Gauss. Bulk Lorentz factor $\Gamma = 2.0$ (corresponding to $P_{||}/P_{\perp} = 14$). We have used $\gamma = 10^6$ and $\gamma = 3.3\times10^5$ for the random Lorentz factor of the TeV emitting electrons.}
%\end{figure}

\label{lastpage}


\begin{thebibliography}{}
\bibitem{}[] Abdo A et al., 2011, ApJ, 727, 129 (Abdo et al 2011a)
\bibitem{}[] Abdo A et al., 2011, ApJ, 736, 131 (Abdo et al 2011b)
\bibitem{}[] Aharonian F. et al., 2007, ApJ, 664, L71
\bibitem{}[] Aharonian F., Timokhin, A. N., Plyasheshnikov, A. V., 2002, A\&A, 384, 834
\bibitem{}[] Albert J. et al., 2007, ApJ, 669, 862
\bibitem{}[] Baker D. N., Borovsky J. E., Benford G., Eilek J. A., 1988, ApJ, 326, 110
\bibitem{}[] Becker P. A., Kafatos M. 1995, ApJ, 453, 83
\bibitem{}[] Begelman M. C., Fabian A. C., Rees M. J., 2008, MNRAS, 384, L19
\bibitem{}[] Blazejowski M. et al., 2005, ApJ, 630, 130
\bibitem{}[] Crusius-W\"atzel A. R., Lesch, H., 1998, A\&A, 338, 399
\bibitem{}[] Eilek J. A., 1979, ApJ, 230, 373
\bibitem{}[] Georganopoulos, M., Kazanas, D., 2003, ApJ, 594, L27
\bibitem{}[] Ghisellini G., Tavecchio F., Bodo G., Celotti A., 2009, MNRAS, 393, L16
\bibitem{}[] Giannios D., Uzdensky D. A., Begelman M. C., 2010, MNRAS, 402, 1649
\bibitem{}[] Giroletti, M. et al., 2004, ApJ, 600, 127
\bibitem{}[] Gopal-Krishna, Goyal A., Joshi S., Karthick C., Sagar R., Wiita P. J., Anupama G. C., Sahu D. K., 2011, MNRAS, 416, 101
\bibitem{}[] Kirk J. G., Mochol I., 2011, ApJ, 729, 104
\bibitem{}[] Krall N. A., Trivelpiece A. W., 1973, Principles of Plasma Physics, McGraw Hill, New York
\bibitem{}[] Krishan V., Wiita P. J., 1994, ApJ, 423, 172
\bibitem{}[] Levinson, A., 2007, ApJ, 671, L29
\bibitem{}[] Lucek S. G., Bell A. R., 2000, MNRAS, 314, 65
\bibitem{}[] Melrose D. B., 1970, Ap\&SS, 6, 321
\bibitem{}[] Noerdlinger P. D., Yui A. K-M., 1969, ApJ, 157, 1147
\bibitem{}[] Paesold G., Benz A. O., 1999, A\&A, 351, 741
\bibitem{}[] Piner, B. G., Pant, N., Edwards, P. G., 2008, ApJ, 678, 64
\bibitem{}[] Shukla, A., Chitnis, V. R., Vishwanath, P. R., Acharya, B. S., Anupama, G. C., Bhattacharjee, P., Britto, R. J., Prabhu, T. P., Saha, L., Singh, B. B., 2012, to appear in A\&A, arXiv:1203.3850
\bibitem{}[] Subramanian, P., Becker, P. A., Kazanas, D. 1999, ApJ, 523, 203
\bibitem{}[] Urry, C. M., Padovani, P., 1991, ApJ, 371, 60
\bibitem{}[] Wentzel D. G., 1968, ApJ, 152, 987
\bibitem{}[] Wentzel D. G., 1969, ApJ, 157, 545
\end{thebibliography}
\end{document}